\documentclass[draft]{agujournal2019}
\usepackage{url} 
\usepackage{lineno}
\usepackage[inline]{trackchanges} 
\usepackage{soul}
\usepackage{amssymb, amsmath, physics}
\draftfalse
\usepackage{xcolor}

\showjournalnamefalse
\journalname{Geophysical Research Letters}

\begin{document}

\title{Critical ocean physics on the edge of AMOC instability}

\authors{Reyk Börner\affil{1,2}, Anna S. von der Heydt\affil{1,2}, Robbin Bastiaansen\affil{1,2,3}, and Henk A. Dijkstra\affil{1,2}}
\affiliation{1}{Institute for Marine and Atmospheric research Utrecht, Department of Physics, Utrecht University, Utrecht, The Netherlands}
\affiliation{2}{Centre for Complex Systems Studies, Department of Physics, Utrecht University, Utrecht, The Netherlands}
\affiliation{3}{Mathematical Institute, Utrecht University, Utrecht, The Netherlands}

\correspondingauthor{Reyk Börner}{r.borner@uu.nl}

\begin{keypoints}
\item We determine an edge state of the Atlantic Meridional Overturning Circulation in the global ocean model POP2 at 1$^\circ$ horizontal resolution.
\item We characterize the physics of the edge state, focusing on the ocean density field and freshwater and heat budgets in the North Atlantic.
\item Our results support using the regionally integrated surface buoyancy flux as a physics-based indicator for the onset of AMOC tipping.
\end{keypoints}

\begin{abstract}
The Atlantic Meridional Overturning Circulation (AMOC) is expected to weaken under global warming, but whether this signifies a transient response or a tipping transition remains unclear. Distinguishing between these cases requires a physical understanding of the boundary separating ocean states leading to AMOC recovery from those leading to collapse. In a global ocean model with 1$^\circ$ resolution, we compute an AMOC edge state -- an unstable saddle equilibrium that anchors this instability boundary. The edge state maintains weak overturning sourced by marginal deep water formation in the Nordic Seas. Analyzing the North Atlantic freshwater and heat balances that govern the edge state dynamics, we show that near the edge state, thermal and haline components of the surface buoyancy flux approximately compensate. Surface buoyancy loss (gain) coincides with AMOC recovery (collapse). Our findings support using the regional net surface buoyancy flux to inform an indicator for the onset of AMOC tipping.
\end{abstract}

\section*{Plain Language Summary} 
The Atlantic Meridional Overturning Circulation (AMOC) carries warm surface water northward and returns cold, deep water southward, shaping the climate of the Northern Hemisphere. Global warming could cause this circulation to weaken or even collapse, but identifying the ocean conditions that separate a temporary weakening from collapse due to crossing a tipping point remains difficult. Using a global ocean model, we identify an unstable circulation state, also called edge state, that lies on the boundary between these two outcomes. We compare its temperature and salinity patterns and exchanges of heat and freshwater with those of strong and collapsed circulation states. At this boundary, the combined effects of surface heating, cooling, and freshwater exchange on seawater density nearly balance across the North Atlantic regions that supply deep waters. Departures toward making surface waters lighter lead to freshwater accumulation and disrupt the ocean pathways that sustain the circulation, causing collapse. Departures in the opposite direction allow recovery. Our results suggest that monitoring these surface exchanges could help identify when the AMOC begins to collapse.

\section{Introduction}

Climate models project a weakening of the Atlantic Meridional Overturning Circulation (AMOC) over the 21st century \cite{weijer_cmip6_2020,ipcc_climate_2023}. Given its major role in the global ocean circulation and northward heat transport, a substantial AMOC decline would likely have large-scale climate impacts \cite{jackson_global_2015,liu_climate_2020,ben-yami_impacts_2024}. However, it remains uncertain whether AMOC weakening under anthropogenic forcing represents a transient response with subsequent recovery or a tipping transition to a persistent collapsed circulation state \cite{bonan_transient_2022,guo_reversible_2026,drijfhout_shutdown_2025}.

The tipping potential of the AMOC is supported by a hierarchy of climate models \cite{dijkstra_transitions_2026}. Quasi-equilibrium hysteresis experiments imposing a slowly varying freshwater input over the North Atlantic consistently demonstrate a multistable AMOC regime in which a strong and a collapsed AMOC state coexist \cite{stommel_thermohaline_1961a,rahmstorf_thermohaline_2005,hawkins_bistability_2011,vanwesten_asymmetry_2023}. A key mechanism underlying this multistability is the salt-advection feedback \cite{stommel_thermohaline_1961a,rahmstorf_freshwater_1996,dijkstra_stability_2005,vanderborght_feedback_2025}, enabling self-sustaining AMOC change once a critical threshold has been crossed. Yet, distinguishing a tipping transition from transient weakening is difficult based on AMOC strength alone \cite{borner_tip_2026}. This distinction matters for understanding the stability and resilience of the AMOC, particularly its nonlinear sensitivity and reversibility under climate change scenarios.

Under quasi-equilibrium freshwater forcing, the tipping threshold is commonly identified as a bifurcation where the current AMOC state loses stability. The scenario of bifurcation-induced tipping \cite{ashwin_tipping_2012a} forms the basis for statistical early-warning indicators based on critical slowing down, as applied to observational AMOC fingerprints \cite{boers_observationbased_2021,ditlevsen_warning_2023}. However, their interpretation is limited for practical and fundamental theoretical reasons \cite{ben-yami_uncertainties_2023,lohmann_predicting_2026}. This calls for indicators that connect directly to the physical processes governing AMOC stability. For example, the sign of the freshwater import via overturning into the Atlantic through its southern boundary, $F_\text{ovS}$, provides insight into whether salt advection acts to stabilize or destabilize the AMOC \cite{rahmstorf_freshwater_1996,devries_atlantic_2005}. Yet, its interpretation as a stability indicator also requires quasi-equilibrium conditions \cite{rahmstorf_freshwater_1996,vanwesten_physicsbased_2025a}.

Under transient forcing, AMOC transitions may occur without crossing a bifurcation, e.g. via rate-induced tipping \cite{wieczorek_rateinduced_2023,stocker_influence_1997,vanwesten_failure_2026,alkhayuon_basin_2019}. In this case, time-dependent forcing drives the ocean out of equilibrium, and the tipping indicators mentioned above are not expected to apply. Instead, \citeA{vanwesten_physicsbased_2025a} proposed the integrated surface buoyancy flux over a North Atlantic region, $B_\text{flux}$, as a tipping onset indicator. A sign change from negative to positive values represents a transition from net surface buoyancy loss to gain, inhibiting the surface-forced supply of dense water sustaining the AMOC \cite{walin_relation_1982,speer_rates_1992}. While this indicator demonstrated skill under quasi-equilibrium, freshwater-pulse, and anthropogenic forcing \cite{vanwesten_physicsbased_2025a}, its robustness and theoretical foundation requires clarification.

Out of equilibrium, the relevant `threshold' is a boundary that separates, for a given fixed forcing, states attracted by the strong versus collapsed AMOC equilibrium. This boundary is a high-dimensional surface in the system's state space, possibly fractal \cite{bodai_rough_2020,mehling_limits_2024}, that generally depends on the forcing conditions. However, the dynamics on this surface typically tend towards one or more saddle equilibria, termed edge states. Although unstable, edge states can play an important role in tipping dynamics and transient behavior \cite{lucarini_edge_2017,feudel_rateinduced_2023,borner_global_2026,lohmann_role_2025,lohmann_predicting_2026}, offering a global stability perspective that can help understand why a weakened AMOC sometimes recovers and sometimes collapses.

Unstable AMOC solutions have been computed using numerical continuation \cite{dijkstra_stability_2005} or via an edge tracking technique that relies solely on forward integration from modified initial conditions. Recent studies used edge tracking to determine an AMOC edge state in conceptual chaotic climate models \cite{mehling_limits_2024,axelsen_finitetime_2024}, in a coarse-resolution global ocean model \cite{lohmann_melancholia_2024} and in an earth system model of intermediate complexity \cite{borner_global_2026}, revealing its potential role in ensemble splitting, long transients and a boundary crisis \cite{borner_global_2026}.

Here, we take a next step by determining an edge state of the AMOC in the ocean component of a Coupled Model Intercomparison Project (CMIP) model. Our study focuses on how the physical processes underlying the $B_\text{flux}$ indicator are related to the edge state. We characterize the physics of the edge state in comparison with the two stable equilibria and quantify the buoyancy balances that maintain it. Our results suggest that a vanishing $B_\text{flux}$ correlates reasonably well with the separating boundary forming the stable manifold of the edge state. By linking dynamical systems ideas with oceanographic analysis, we hope to improve our understanding of the AMOC under present and future forcing.

\section{Methods}
\subsection{Ocean Model}
We use the Parallel Ocean Program v2 (POP2), the ocean component of the Community Earth System Model (CESM), with a nominal horizontal resolution of $1^\circ$ and 40 vertical layers \cite{danabasoglu_ccsm4_2012,smith_parallel_2010}.
The ocean-only configuration is forced with a repeating seasonal cycle of a prescribed atmospheric climatology using mixed surface boundary conditions (see Supporting Text S1). Despite precipitation, river runoff, downward shortwave radiation and atmospheric temperatures being fixed, the net surface freshwater and heat fluxes can vary because the simulated sea surface temperature (SST) controls evaporation and the latent, sensible, and net longwave heat fluxes.
The prescribed, time-independent sea ice distribution roughly corresponds to boreal winter conditions (Supporting Fig. S1). 

This idealized setup neglects ocean-atmosphere-ice interactions, isolating the ocean processes governing AMOC instability. The practical advantage is that simulated trajectories exhibit negligible high-frequency variability due to the lack of chaotic atmospheric and sea-ice dynamics. Equilibrium states approximately become stationary fixed points. This simplifies the edge state computation, making it practically feasible in this high-dimensional model.

The AMOC stability properties of our POP2 configuration have been studied as a function of a freshwater hosing flux $F_H$ \cite[]{vanwesten_collapse_2025}. A quasi-equilibrium hysteresis experiment revealed a bistable AMOC regime in which a strong (`On') and collapsed (`Off') AMOC steady state coexist (Fig. \ref{fig:1}a). 
Here, we do not apply any hosing ($F_H = 0$), placing us roughly in the center of the bistable regime.

\subsection{Edge Tracking}
An edge state is an unstable saddle equilibrium whose stable manifold locally corresponds to the basin boundary separating the basins of attraction of the coexisting stable states (Fig. \ref{fig:1}b). Its unstable manifold traces the relaxation pathways from the edge state to the On and Off state, respectively. Trajectories near the basin boundary initially tend towards the edge state before diverging to one of the attractors. By counteracting this divergence, a simulation technique known as the edge tracking algorithm allows to trace a path on the basin boundary that converges to the edge state \cite{battelino_multiple_1988,skufca_edge_2006,lucarini_edge_2017}.

Starting from two ocean states that evolve towards different attractors (thus lying on opposite sides of the basin boundary), we obtain an intermediate initial condition by averaging the two model restart files. Repeating this bisection yields a pair of similar ocean states that still evolve to opposite attractors (Fig. \ref{fig:1}b). The pair is integrated forward in time. When the separation between the trajectories becomes too large, the bisection step is repeated and the new resulting pair is evolved in time. This iterative procedure produces trajectory segments that `shadow' a path on the basin boundary on either side of it. From these segments, we construct an \textit{edge trajectory}, i.e. a pseudo-trajectory that approximates a trajectory on the basin boundary. After a transient, the edge trajectory is expected to converge to an edge state and approximately sample the edge state dynamics. 
The quality of the estimate depends on the number and frequency of bisections (see \citeA{borner_climate_2025} for a detailed description).

Edge tracking has been applied to investigate the edge of chaos in turbulent flows \cite{skufca_edge_2006,schneider_laminarturbulent_2008,brynjell-rahkola_chaotic_2024}, global climate instability \cite{lucarini_edge_2017,lucarini_global_2020}, and AMOC edge states \cite{mehling_limits_2024,axelsen_finitetime_2024,lohmann_melancholia_2024,borner_global_2026}.

\begin{figure}
\noindent\includegraphics[width=\textwidth]{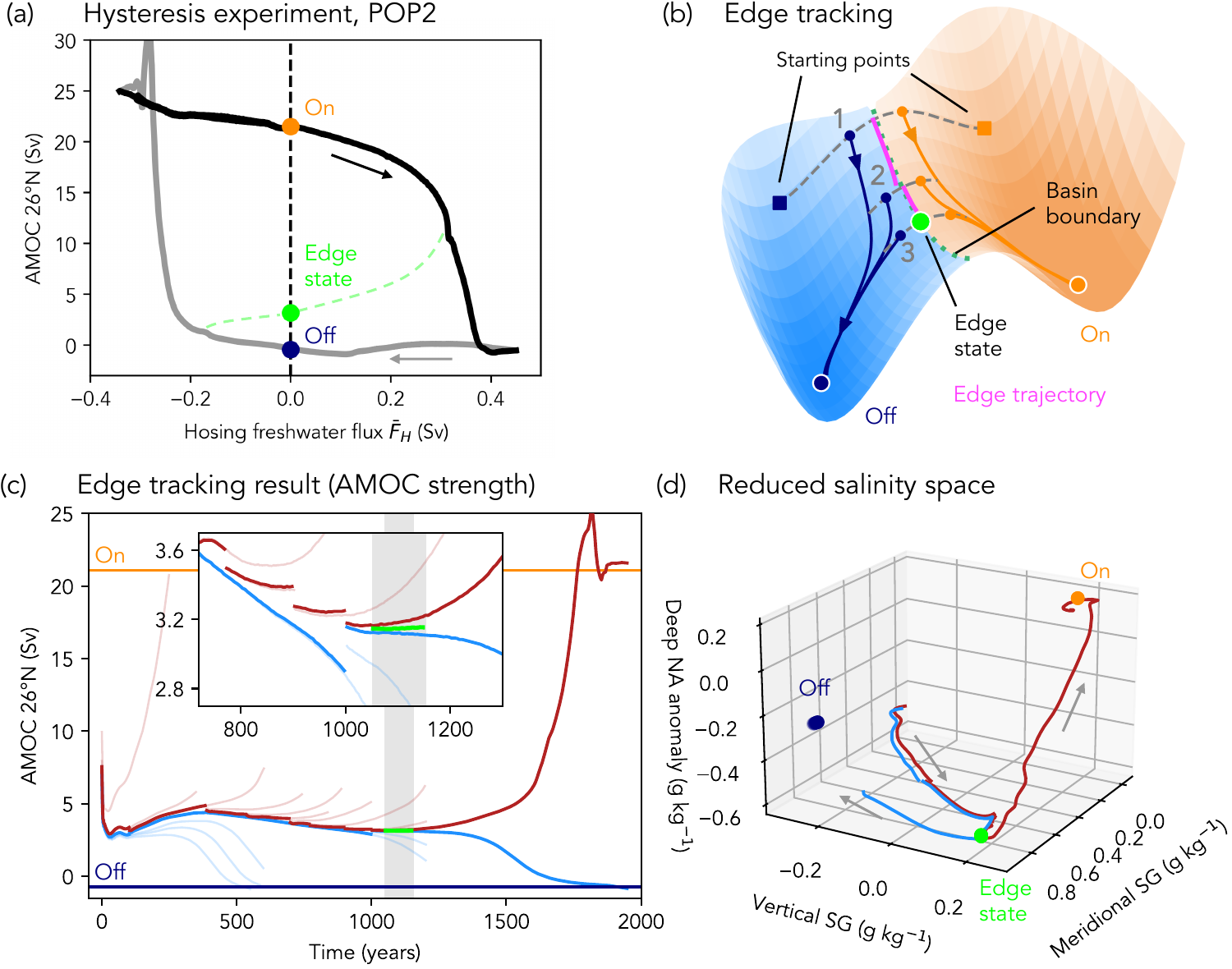}
\caption{(a) Quasi-equilibrium hosing experiment \cite{vanwesten_collapse_2025}, demonstrating a bistable AMOC regime. AMOC strength is measured at 26$^\circ$N and 1000\,m depth. The On (orange), Off (blue) and edge (green) states are indicated at $F_H = 0$. The edge state lies on the unstable branch (non-computed, hypothetical green dashed line). (b) Schematic of the edge tracking algorithm in the stability landscape of a bistable system, colored by basin of attraction. Edge tracking starts from a pair of states (squares), with bisection steps indicated (gray), converging to the edge state. (c) Edge tracking trajectory segments shown as AMOC timeseries, leading to On (red) or Off (blue). The shaded 100-year interval is used to estimate the edge state (green). (d) like (c) but projected in the reduced salinity space (SG = salinity gradient, NA = North Atlantic; see Supporting Text S3).}
\label{fig:1}
\end{figure}

\section{Results}

\subsection{Edge State Determination}
We begin edge tracking from the On and Off attractors, obtained by branching off 50-year simulations from the quasi-equilibrium hysteresis run at $F_H=0$ (Fig. \ref{fig:1}a). These states have an AMOC strength of around 21~Sv (On) and 0~Sv (Off), measured at 26$^\circ$N and 1000 m depth (1 Sv $= 10^6$ m$^3$ s$^{-1}$). The first bisection step (details in Supporting Text S2) yields two initial conditions with AMOC strengths of around 7.6 and 7.2~Sv, respectively. Their trajectories rapidly decline to below 4~Sv within 5 years but eventually evolve to different attractors (Fig. \ref{fig:1}c). 
The subsequent sequence of edge tracking iterations produces an edge trajectory that wavers between 3 and 4.5~Sv and, after around 1050 years, seems to stabilize around 3.2~Sv.

The first 1000 years of the edge trajectory represent a curve on the basin boundary. We estimate the edge state from the edge trajectory segment between years 1050 and 1150, averaging in time over that 100-year period. To approximate the edge state's unstable manifold, we continue the trajectory pair of the final edge tracking iteration for another 750 years. These trajectories indeed approach the On and Off state, respectively (Fig \ref{fig:1}c).

The edge state dynamics can be visualized in a reduced state space projecting the Atlantic salinity field (Fig. \ref{fig:1}d, Supporting Text S3) \cite{borner_global_2026}. The edge state does not simply lie `in between' the On and Off states: the edge trajectory departs substantially from the bisection line connecting the attractors, and the edge state is characterized by distinct Atlantic salinity properties.

\subsection{Edge State Characterization}
\begin{figure}
\noindent\includegraphics[width=\textwidth]{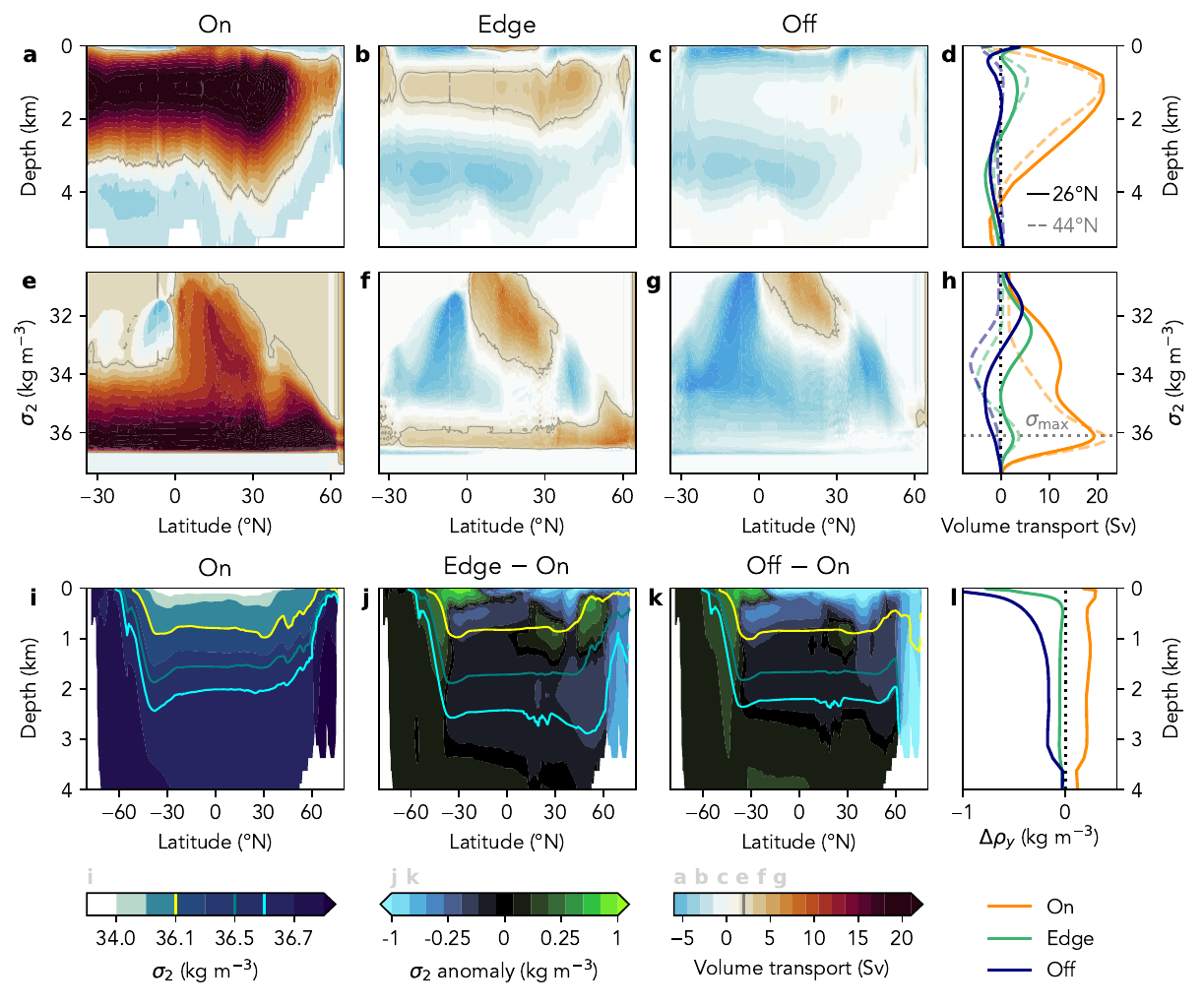}
\caption{(a-c) AMOC streamfunction in depth and (e-g) density ($\sigma_2$) space of the On, edge and Off states, respectively. (d) Profiles of overturning strength at 26$^\circ$N (solid) and 44$^\circ$N (dashed) in depth and (h) density space, indicating $\sigma_2^\text{max}$ (dotted gray). (i-k) Potential density ($\sigma_2$) cross-sections in the Atlantic sector (bounded by $70^\circ$W and 20$^\circ$E south of 34$^\circ$S), indicating isopycnals at $\sigma_2^\text{max}=36.1$~kg\,m$^{-3}$ (yellow), 36.5~kg\,m$^{-3}$ (teal) and 36.6~kg\,m$^{-3}$ (cyan). Colored shading in (j) and (k) shows anomalies relative to (i). (l) Profiles of $\Delta \rho_y = \rho_N - \rho_S$ ($\sigma_0$ coordinates).}
\label{fig:2}
\end{figure}

\begin{figure}
\noindent\includegraphics[width=\textwidth]{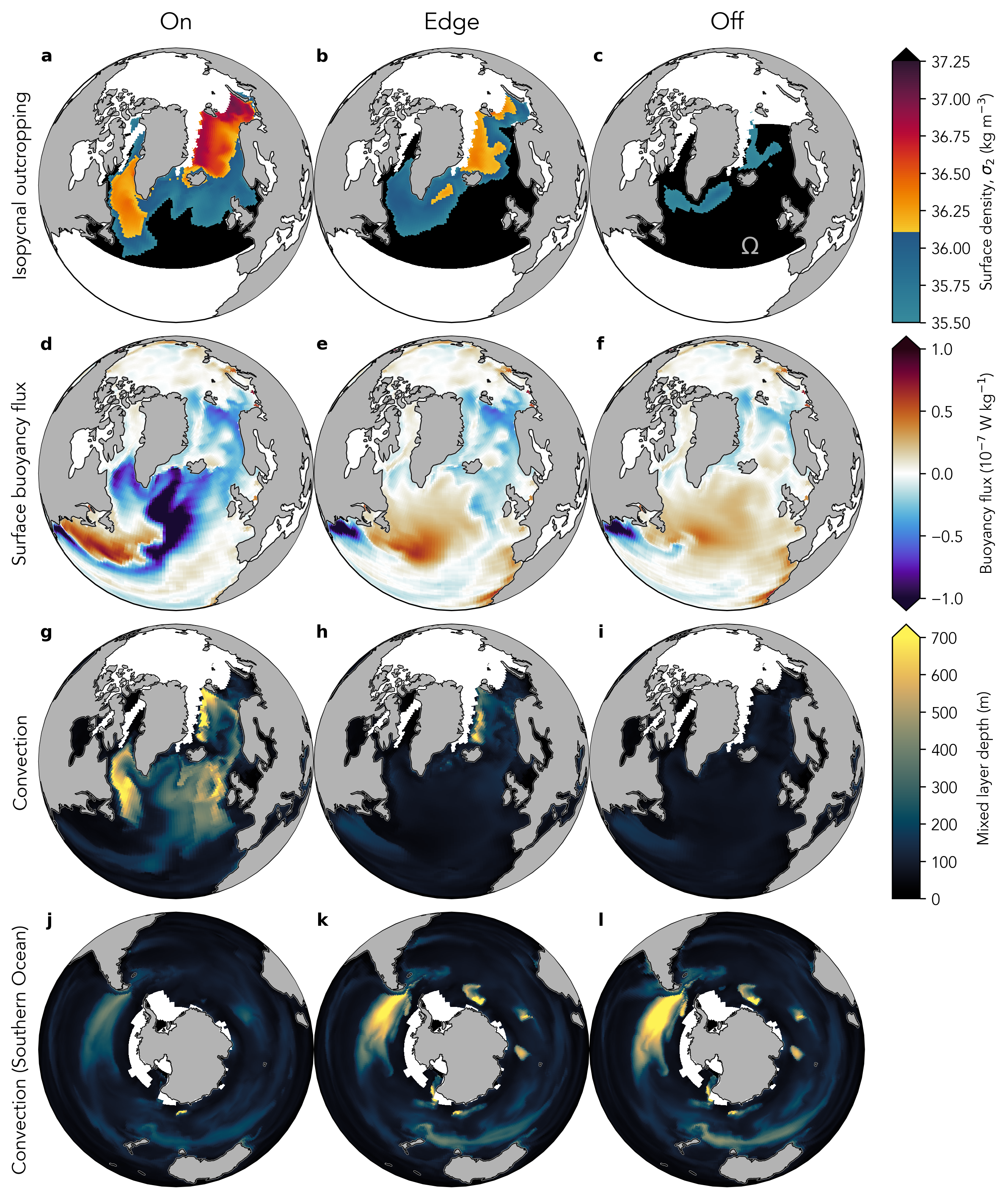}
\caption{(a-c) Surface density in $\sigma_2$ coordinates, showing only densities above 35.5~kg\,m$^{-3}$ (blue shading) and above $\sigma_2^\text{max}$ (orange-red). The black area marks $\Omega$. (d-f) Net surface buoyancy flux (into the ocean). (g-i) Winter mean mixed layer depth in the North Atlantic (months JFM) and (j-l) Southern Ocean (months JJA). Sea ice-covered regions are white.}
\label{fig:3}
\end{figure}

We compare the steady-state properties of the AMOC On, Off and edge states (Figs. \ref{fig:2} and \ref{fig:3}) before analyzing the dynamical processes underlying the edge state in section \ref{sec:physics}.  
The edge state sustains a weak Atlantic meridional overturning cell with a depth-space maximum of around 5.6~Sv at 44$^\circ$N and 700~m depth (Fig. \ref{fig:2}b). The streamfunction in density space reveals that this cell is connected up to high latitudes ($>65^\circ$N, Fig. \ref{fig:2}f). In both the On and edge states, the maximum overturning occurs at a potential density of $\sigma_2^\text{max} \approx 36.1$~kg\,m$^{-3}$ (Fig. \ref{fig:2}h). The edge state also features a strengthened bottom water cell at depths below 2500~m. In the Off state (Fig. \ref{fig:2}c,g), the bottom cell remains but the AMOC overturning cell disappears, with no northward overturning transport at any density level at 44$^\circ$N (Fig. \ref{fig:2}h). 

%
From a water-mass transformation perspective \cite{groeskamp_water_2019}, the AMOC is sustained by adiabatic pathways along isopycnals that outcrop both in the Southern Ocean and in the North Atlantic \cite{wolfe_adiabatic_2011,nikurashin_theory_2011}. The closure of these pathways due to a loss of northern outcropping constitutes an AMOC collapse.
In the On state, isopycnals with potential densities around $\sigma_2^\text{max}$ indeed outcrop on both hemispheres (Fig. \ref{fig:2}i). Northward, they surface in two main regions located in the Labrador and Nordic Seas (Fig. \ref{fig:3}a). By constrast, the Off state exhibits a complete loss of outcropping above $\sigma_2^\text{max}$ in the North Atlantic (Fig. \ref{fig:3}c). The edge state maintains a reduced outcropping region in the Nordic Seas while outcropping ceases in the Labrador Sea (Fig. \ref{fig:3}b).

The zonally averaged potential density field in the Atlantic sector provides a subsurface view of these differences (Fig. \ref{fig:2}i-k). In the deep Atlantic north of 60$^\circ$N, the On state contains heavy water masses, with potential densities exceeding those in the deep Southern Ocean, whereas these deep North Atlantic waters are substantially lighter in the Off and edge states. Southern Ocean densities are overall higher in the Off and edge states (relative to the On state), especially towards the surface around 34$^\circ$S. This could be related to enhanced Southern Ocean convection (Fig. \ref{fig:3}k,l). The $\sigma_2^\text{max}$ isopycnal mainly differs in the subpolar North Atlantic, where it grazes the surface in the edge state but remains below 500 m depth in the Off state. Furthermore, isopycnals in the central interior Atlantic (around 36.6 kg m$^{-3}$) are shifted downwards in the edge state (Fig. \ref{fig:2}j), reflecting that the edge state has the least stratified ocean compared with the On and Off states (Supporting Fig. S4).

\citeA{vanderborght_feedback_2025} showed in CESM that, in equilibrium, the AMOC strength can be reconstructed from the meridional density gradient $\Delta \rho_y = \rho_N - \rho_S$, where $\rho_N(z)$ and $\rho_S(z)$ are the horizontally-averaged potential density profiles over a northern and southern latitude range in the Atlantic sector, respectively. When choosing these latitude ranges to reflect the On state locations of isopycnal outcropping (34-65$^\circ$S, 50-80$^\circ$N), we find that $\Delta\rho_y$ of the edge state is close to zero across the water column, except near the surface (Fig. \ref{fig:2}l). By contrast, the On (Off) state has a positive (negative) $\Delta\rho_y$ across the water column.

Differences in the density field across the three states coincide with a spatial reorganization of the surface buoyancy flux (Fig.~\ref{fig:3}d--f). The On state exhibits strong buoyancy loss over much of the subpolar North Atlantic, particularly in the Labrador and Irminger Seas and along the North Atlantic Current (Fig.~\ref{fig:3}d). In the edge state, a dipole pattern appears with gain over the subpolar gyre and loss in the Nordic Seas (Fig.~\ref{fig:3}e). Thus, the edge state retains regions of surface buoyancy loss where dense isopycnals still outcrop. In the Off state, buoyancy loss in the Nordic Seas is further reduced (Fig.~\ref{fig:3}f).

As a measure of deep convection, we consider the winter-mean mixed layer depth (Fig.~\ref{fig:3}g--l). In the On state, deep mixed layers occur in the Labrador Sea, south of Iceland, and in the Nordic Seas, locally exceeding 700\,m (Fig.~\ref{fig:3}g). In the edge state, most of these deep convection regions disappear, except for a region in the Nordic Seas (Fig.~\ref{fig:3}h), consistent with the restricted isopycnal outcropping of the edge state. In the Off state, mixed layers are shallow everywhere in the North Atlantic (Fig.~\ref{fig:3}i), indicating a shutdown of deep convection.
In the Southern Ocean, regions of deepened mixed layers suggest an activation of deep convection in the edge and Off states compared to the On state (Fig.~\ref{fig:3}j--l). This is consistent with the strengthened bottom water cell of these states (Fig.~\ref{fig:2}b,c). The edge state combines a Southern Ocean convection pattern resembling that of the Off state with localized North Atlantic convection that sustains a weak AMOC.

\subsection{Edge State Physics \label{sec:physics}}
\begin{figure}
\noindent\includegraphics[width=\textwidth]{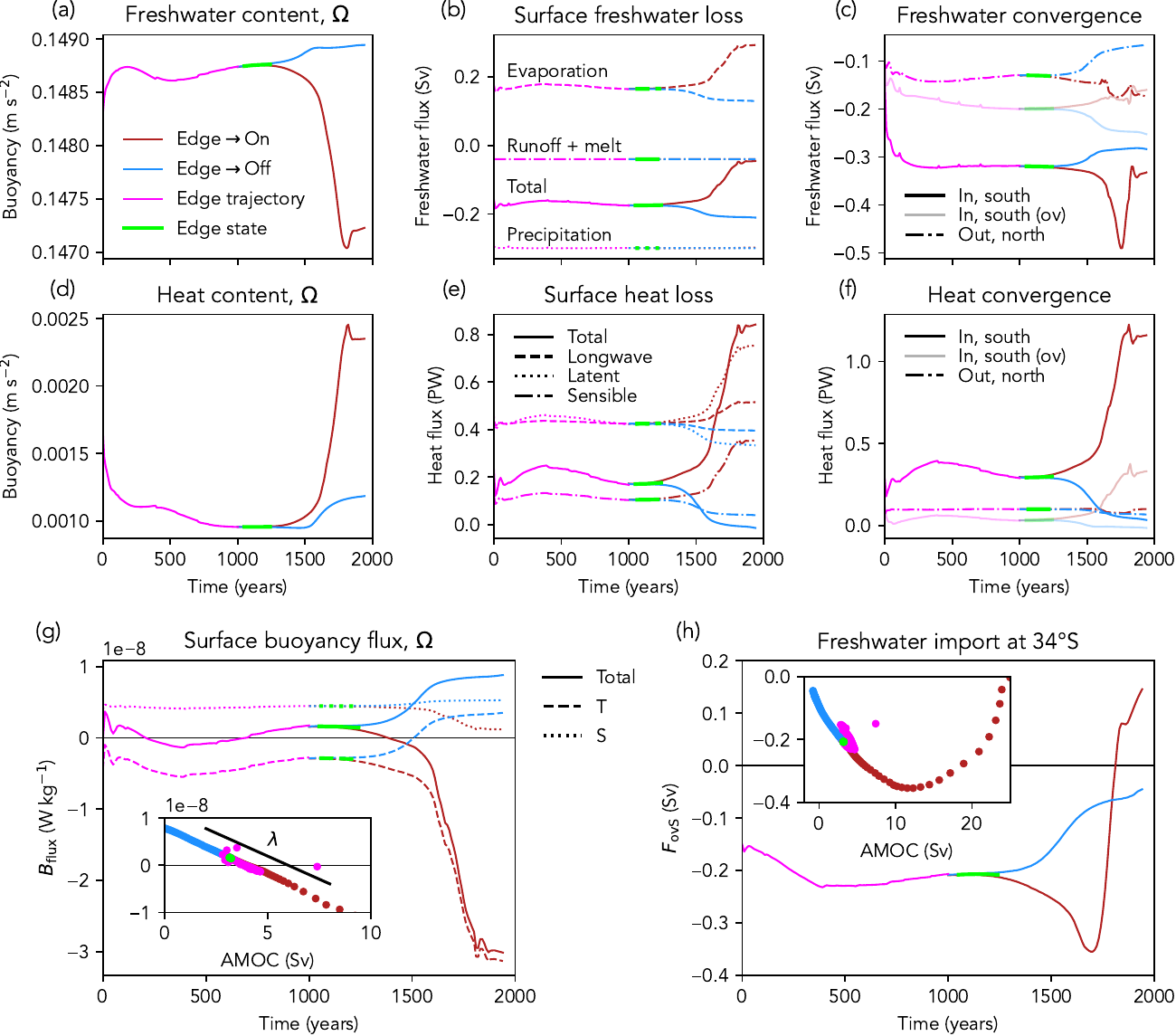}
\caption{Freshwater and heat budgets along the edge trajectory (magenta) and relaxation paths from the edge state (green) towards On (red) and Off (blue). (a) Freshwater and (d) heat concent in $\Omega$ in buoyancy units (Supporting Text S3). (b) Surface freshwater and (e) heat flux decomposition (positive leaving the ocean). (c) Meridional freshwater and (f) heat convergence in $\Omega$, separated by southern incoming (total, solid; overturning component, faint) and northern outgoing (total, dash-dotted) transport. (g) $B_\text{flux}$ integrated over $\Omega$, decomposed into thermal (T) and haline (S) contributions. Inset: $B_\text{flux}$ vs. AMOC strength; slope $\lambda$ from least-squares fit. (b) $F_\text{ovS}$ vs. time and (inset) AMOC strength.}
\label{fig:4}
\end{figure}

To investigate the physical balances governing the edge state dynamics, we decompose the freshwater and heat budgets of the North Atlantic volume $\Omega$ (Fig.~\ref{fig:3}c, Supporting Fig. S1) into surface and lateral fluxes. Bounded in the south by about $44^\circ$N, the northern boundary of $\Omega$ follows the model grid between $71^\circ$N and $77^\circ$N. This choice, rather than the 40-65$^\circ$N latitude band used by \citeA{vanwesten_physicsbased_2025a}, includes more of the isopycnal outcropping regions identified above, covering most of the Labrador, Irminger and Nordic Seas. 
%
%

We compute budgets over time along the edge trajectory and the relaxation trajectories leading from the edge state to the On and Off state, respectively (Fig. \ref{fig:4}). 
The freshwater content in $\Omega$ varies slightly along the edge trajectory but lies between that of the On and Off states (Fig. \ref{fig:4}a). AMOC collapse coincides with freshwater accumulation, driven by reduced evaporation (Fig. \ref{fig:4}b) and reduced southern salt import, while the latter is overcompensated by reduced northern salinity export (Fig. \ref{fig:4}c). AMOC recovery is associated with strong salinification, where a simultaneous horizontal salt import and enhanced evaporation contribute about equally. The southern salt import is driven mainly by the gyre component, whereas the overturning component slightly opposes the change (Fig. \ref{fig:4}c, Supporting Fig. S3).

The heat budget tendencies are qualitatively different. Convergence to the edge state coincides with heat decay in $\Omega$, and the edge state has the lowest heat content of all states (Fig. \ref{fig:4}d). As expected, the heat convergence into $\Omega$ is dominated by heat import from the south (Fig. \ref{fig:4}f) and correlates well with the AMOC strength. In equilibrium, this heat input is compensated by surface heat loss and a small loss through the northern boundary. Since atmospheric conditions are fixed, the heat fluxes mainly respond to SST changes. These dynamic heat loss components all decrease towards AMOC collapse and increase towards recovery. Along the edge trajectory, the latent and net longwave heat flux contribute equally. The Off state features little heat exchange.

The edge state combines a substantial surface freshwater input with much weaker net surface heat loss than the On state. Whether the surface forcing increases or decreases ocean density depends on the competition of thermal and freshwater effects, measured by the surface buoyancy flux integrated over $\Omega$, $B_\text{flux}$ (Fig. \ref{fig:4}g). 

Along the edge trajectory, its positive freshwater contribution is largely compensated by its negative thermal contribution, leaving a small residual around $B_{\mathrm{flux}}=0$. AMOC recovery is accompanied by increasing net surface buoyancy loss, attributed mostly to the thermal contribution, supported by a decreasing haline contribution. During AMOC collapse, the thermal contribution switches sign from negative to positive while the haline contribution remains positive, causing net surface buoyancy gain. 

$B_{\mathrm{flux}}$ is linearly correlated to the AMOC strength both when converging to and diverging from the edge state ($R^2=0.993$, slope $\lambda \approx -1.95\times 10^{-9}$ W kg$^{-1}$ Sv$^{-1}$, inset Fig. \ref{fig:4}g). Thus, the value of $B_\text{flux}$ at the edge state separates AMOC states during recovery from those during collapse. 
Contrarily, $F_\text{ovS}$ is not well suited as an indicator of the basin boundary (Fig. \ref{fig:4}h). Although $F_\text{ovS}$ depends roughly quadratically on the AMOC strength, the minimum is reached during recovery. This contrasts with the finding by \citeA{vanwesten_physicsbased_2024} that a minimum in $F_\text{ovS} >0$ precedes the tipping onset. Nonetheless, the edge state value of $F_\text{ovS} \approx -0.2$ Sv is still negative and lower than both the On and Off states. This suggests the presence of a state space region at $F_H=0$ where the salt-advection feedback is destabilizing, even though $F_\text{ovS} >0$ in the On state.

\section{Discussion}
Understanding the AMOC's response to transient forcing requires knowledge of the system's global stability properties \cite{margazoglouDynamicalLandscapeMultistability2021,lohmann_predicting_2026,borner_global_2026}. Edge states often mediate tipping dynamics and provide information about the basin boundary \cite{feudel_rateinduced_2023}. Recent conceptual work has explored the link between edge states and early warning indicators of AMOC tipping \cite{lohmann_role_2025,chapman_quantifying_2025,ritchie_evaluating_2026}. Here, we bridge the gap towards tipping indicators for comprehensive climate models by explicitly computing an AMOC edge state in a CMIP-class global ocean model.

The edge state reveals critical ocean conditions that separate AMOC recovery from collapse: localized surface buoyancy loss enables limited isopycnal outcropping and deep convection in the Nordic Seas, supporting a weak overturning cell. The importance of Nordic Seas convection for AMOC stability has been suggested in previous studies \cite{roewer_nordic_2026,arthun_surfaceforced_2023}. 

The integrated surface buoyancy flux in the North Atlantic, $B_\text{flux}$, remains close to zero along the edge trajectory, although our results do not establish $B_\text{flux}=0$ as an exact property of the basin boundary. The value of $B_\text{flux}$ is also sensitive to the choice of integration region. Nonetheless, AMOC recovery (collapse) from the edge state is accompanied by increasingly negative (positive) $B_\text{flux}$ values. Our results therefore offer a physical and dynamical interpretation of the sign change (negative to positive) of $B_\text{flux}$ proposed as an indicator for the onset of AMOC tipping \citeA{vanwesten_physicsbased_2025a}. This indicator captures surface water mass transformation processes that cannot be inferred from AMOC strength alone. Although we find a linear correlation between AMOC strength and $B_\text{flux}$ at fixed forcing, transiently forced simulations show that no particular AMOC strength defines a unique threshold across forcing pathways. Potentially, $B_\text{flux}$ offers a more forcing-independent tipping diagnostic. Still, a transient positive $B_\text{flux}$ need not imply irreversible collapse if transport compensates the surface flux or forcing changes reverse the sign of $B_\text{flux}$ \cite{vanwesten_physicsbased_2025a}.
The quasi-equilibrium finding by \citeA{vanwesten_physicsbased_2024}, that a minimum in $F_\text{ovs}$ precedes the onset of AMOC tipping, is opposed by our results under transient ocean dynamics. Both results could be reconciled if the edge state approaches the $F_\text{ovs}$ minimum in Fig. \ref{fig:4}h as the hosing flux approaches the bifurcation, which remains to be tested.

Rather than simply being intermediate between the On and Off attractors, the edge state is globally less stratified than the attractors, consistent with the previous finding that the edge state has an elevated dynamic enthalpy \cite{lohmann_melancholia_2024} or oceanic center of mass \cite{borner_global_2026}. The AMOC edge states found in VEROS \cite{lohmann_melancholia_2024}, PlaSim-LSG \cite{borner_global_2026} and here in POP2 also agree in having a colder and fresher deep North Atlantic compared to the attractors. 
By prescribing the atmospheric state and sea ice distribution, our modeling setup omits ocean-atmosphere-ice interactions. Future work is needed to understand how the edge state depends on these additional feedbacks and on forcing parameters such as the hosing flux $F_H$ or atmospheric CO$_2$. While POP2 and VEROS differ in model fidelity and resolution, they are both ocean-only setups, yielding an edge state with a relatively steady AMOC strength. Contrarily, the edge state found in the coupled model PlaSim-LSG exhibits unstable centennial oscillations attributed to sea-ice and atmospheric feedbacks \cite{borner_global_2026}. An investigation of what causes these inter-model differences is left for future work.

This paper contributes an oceanographic perspective to the study of AMOC edge states, complementing recent work on detecting the edge state via its spatio-temporal correlation pattern \cite{moinat_climate_2026} and selecting observables for early warning signals based on edge state information \cite{lohmann_role_2025}. These bridges between dynamical systems theory and climate modelling could help fill critical knowledge gaps in the assessment of AMOC tipping risk under global warming.

\section*{Open Research Section}
Diagnostics were computed using the \texttt{oceanalysis} Python package (\url{https://github.com/reykboerner/oceanalysis}). Processed POP2 model output used to create the figures is archived on Zenodo under the DOI 10.5281/zenodo.24429867 \cite{borner_amoc_2026}.

\section*{Conflict of Interest disclosure}
The authors declare there are no conflicts of interest for this manuscript.

\acknowledgments
This is ClimTip contribution XXX; the ClimTip project has received funding from the European Union’s 
Horizon Europe research and innovation programme under grant agreement No. 101137601. H.A.D. 
was also partially funded by the European Research Council through the ERC-AdG project TAOC 
(project 101055096), and A.S.v.d.H. by the Dutch Research Council through the VICI project VI.C.202.081.
The model simulation and the postprocessing was conducted 
on the Dutch National Supercomputer Snellius, NWO-SURF project 2025.038. 

%
%

\bibliography{mpop}

\renewcommand{\thesection}{S\arabic{section}}
\renewcommand{\thefigure}{S\arabic{figure}}
\renewcommand{\thetable}{S\arabic{table}}
\setcounter{section}{0}
\setcounter{figure}{0}
\setcounter{table}{0}

\newpage
\begin{center}
{\Large Supporting Information}    
\end{center}

\section{POP2 ocean model details}
The model setup used here has been used previously by \citeA{vanwesten_collapse_2025} and \citeA{boot_effect_2026}.

\paragraph{Model grid.} Our ocean-only configuration of POP2 uses a curvilinear dipole grid (``gx1v6'') with the North Pole centered over Greenland (see Fig. \ref{fig:1} for the model geography). The nominal horizontal resolution of 1$^\circ$ amounts to about 50 km locally in the Labrador Sea. Vertical levels have thicknesses ranging from 10 m at the surface to 250 m at depth.

\paragraph{Surface forcing.} The prescribed surface boundary conditions follow a repeating seasonal cycle that is identical every year, based on Coordinated Ocean Reference Experiment (CORE) data \cite{large_diurnal_2004}. Heat fluxes are computed via bulk formulae using prescribed near-surface atmospheric temperatures. Precipitation is prescribed, and river runoff is based on observations of 46 major rivers \cite{fekete_global_1999}. Wind stress is computed offline based on \citeA{hurrell_new_2008}. The sea ice distribution is constant, prescribed based on the $-1.8^\circ$C isoline of the SST climatology (Fig. \ref{fig:1}). In ice-covered regions, the downward shortwave radiation vanishes and surface heat and radiative exchange with the atmosphere is suppressed.

\paragraph{Total salinity.}
POP2 is volume-conserving. Surface freshwater fluxes (including the hosing flux $\bar F_H$) are compensated to integrate approximately to zero, such that the volume-integrated salinity is conserved (up to a small residual). However, with increasing simulation time the model exhibits a small salinity drift ($-1.1\times 10^{-4}$ g kg$^{-1}$ yr$^{-1}$), slowly accumulating additional freshwater. Since the simulations presented here are branched off at different times from the quasi-equilibrium hysteresis experiment (Fig. 1 in main text), and restart files are interpolated, ocean states can have a different global mean salinity by up to 0.07 g kg$^{-1}$ (about 0.2\%). This has a minor effect on our freshwater budget analysis.

\section{Edge tracking details}
Full AMOC trajectories for all edge tracking simulations are shown in Fig. \ref{fig:s2}. In total, we ran 7 iterations, with the number of bisections for each iteration listed in Table \ref{tab:s1}. The number and timing of re-bisections was chosen manually to save computation cost. Table \ref{tab:s1} also shows the trajectories and time segments used to construct the edge trajectory $x_e(t)$ (iterations 0 to 6), defined as
\begin{align}
    x_e(t) = \frac{1}{2} (x^\uparrow(t) + x^\downarrow(t)) \,,
\end{align}
where $t \in [0,1002]$ years and $x^\uparrow$ and $x^\downarrow$ are the concatenated segments of trajectories leading to the On and Off state, respectively.

Interpolated restart files were obtained by averaging two given restart files in all ocean variables.

\begin{table}[ht]
\centering
\begin{tabular}{|l|llllllll|}
\hline
Iteration & 0 & 1 & 2 & 3 & 4 & 5 & 6 & 7\\
\hline
Bisections & 6 & 2 & 2 & 1 & 3 & 4 & 2 & 4 \\
$x^\uparrow$ & it000\_b5 & it001\_b1 & it002\_b2 & it003\_b1 & it004\_b3 & it005\_b4 & it006\_b1 & it007\_b4 \\
$x^\downarrow$ & it000\_b6 & it001\_b2 & it001\_b2 & it001\_b2 & it001\_b2 & it001\_b2 & it001\_b2 & it007\_b2 \\
Years & 0--102 & 103--388 & 389--538 & 539--698 & 699--772 & 773--902 & 903--1002 & 1003--1953 \\
\hline
\end{tabular}
\caption{Edge tracking iterations (see also Fig. \ref{fig:s2}) and number of bisections per iteration. The notation \texttt{itX\_bY} specifies the run launched from the restart file of iteration X and bisection Y. Years indicate the time segments used to construct the edge trajectory.}
\label{tab:s1}
\end{table}

\section{Diagnostics}
Diagnostics are computed using the \texttt{oceanalysis} Python package (\url{https://github.com/reykboerner/oceanalysis}). 
\subsection{Reduced salinity space}
The reduced salinity space is spanned by the following three coordinates \cite{borner_global_2026}:
\begin{itemize}
    \item The meridional salinity gradient (SG) in the Atlantic, measured as the mean salinity difference between 0-20$^\circ$N and 40-80$^\circ$N in the top 1000 m (omitting the top 100 m);
    \item The vertical SG in the North Atlantic, defined as the mean salinity difference between the depths 100-1000 m and 1000-3000 m at 46-66$^\circ$N;
    \item The deep North Atlantic salinity anomaly, defined as the mean salinity anomaly relative to 35 g kg$^{-1}$ in the Atlantic basin north of 50$^\circ$N and below 1000 m depth.
\end{itemize}
This projection is derived from an EOF analysis of edge state dynamics in the PlaSim-LSG model.

\subsection{AMOC streamfunction}
The AMOC streamfunction is computed in depth space as
\begin{align}
    \Psi_z(y, z, t) = - \int_{z_0}^z \int_{x_W}^{x_E} v(x, y, z', t)\, \text{d}x \text{d}z' \,,
\end{align}
and in density space as
\begin{align}
    \Psi_\sigma(y, \sigma, t) = \int_{\sigma_\text{min}}^\sigma \int_{x_W}^{x_E} v(x, y, \sigma', t)\, \text{d}x \text{d}\sigma' \,,
\end{align}
where $v$ is the meridional (northward) velocity at longitude $x$, latitude $y$ and depth $z$ (positive downwards, $z=0$ at the sea surface). Here $z_0$ is the depth of the sea floor and $x_{W,E}$ are the western and eastern boundaries of the Atlantic basin, respectively. Potential density $\sigma$ is given in $\sigma_2$ coordinates; $\sigma_\text{min}$ is the minimum of the chosen density range.

The AMOC strength is defined as $\Psi_z(y=26^\circ\text{N}, \ z=1000\text{ m}, \ t)$.

\subsection{Surface buoyancy flux}
The net surface buoyancy flux, decomposed into thermal (T) and freshwater (S) contributions, is given by
\begin{align}
    B = B_T + B_S = \frac{g \alpha}{\rho_0 c_p} Q_T + \frac{g \beta S}{\rho_0} Q_S \,,
\end{align}
where $Q_T$ and $Q_S$ denote the net surface heat and freshwater flux, respectively. Furthermore, $g$ denotes gravitational acceleration, $c_p$ is the specific heat capacity, $S$ is the sea surface salinity, $\rho_0=1027$ kg m$^{-3}$ is a reference density and $\alpha$ and $\beta$ represent the thermal and haline expansion coefficients, respectively.

\subsection{Heat and freshwater budget}
The heat and freshwater content in an ocean volume $\Omega$, expressed in units of bouyancy (m s$^{-2}$), is given by
\begin{align}
    F_\Omega = - \frac{g}{V}\iiint_\Omega \beta (S-S_0)\, \mathrm{d}V \,, \qquad H_\Omega = \frac{g}{V}\iiint_\Omega \alpha (\theta-\theta_0)\, \mathrm{d}V \,,
\end{align}
where $V = \iiint_\Omega \text{d}V$. The absolute salinity $S$ and conservative temperature $\theta$ have reference values $S_0 = 35$ g kg$^{-1}$ and $\theta_0=0^\circ$C. 

The meridional freshwater transport across a section at latitude $y$,
\begin{align}
    F_\text{tot} (y) = - \frac{1}{S_0} \int_{z_0}^{0}\int_{x_W}^{x_E} vS \mathrm{d}x\,\mathrm{d}z \,,
\end{align}
is decomposed into overturning (ov), gyre (az) and barotropic (bt) components,
\begin{align}
F_{\mathrm{ov}}(y) &= -\frac{1}{S_0}\int_{z_0}^{0}\left[\int_{x_W}^{x_E} v^{*}\,\mathrm{d}x\right]\left[\langle S\rangle - S_0\right]\mathrm{d}z, \\[6pt]
F_{\mathrm{az}}(y) &= -\frac{1}{S_0}\int_{z_0}^{0}\int_{x_W}^{x_E} v'\,S'\,\mathrm{d}x\,\mathrm{d}z, \\[6pt]
F_{\mathrm{bt}}(y) &= -\frac{\overline{S}-S_0}{S_0}\int_{z_0}^{0}\int_{x_W}^{x_E} v\mathrm{d}x\,\mathrm{d}z.
\end{align}
Here $\langle \cdot \rangle$ denotes the zonal mean, $v^*$ and $v'$ are deviations from the section and zonal mean, respectively, and $\overline{S}$ denotes the section (longitude-depth) mean salinity.

The meridional heat transport, 
\begin{align}
    H_\text{tot}(y) = c_p \int_{z_0}^{0}\int_{x_W}^{x_E} v \theta\rho \, \mathrm{d}x\,\mathrm{d}z
\end{align}
can be decomposed analogously. Here $\rho$ denotes in-situ density.

In this work, we did not compute transports across a constant latitude but across a line of curvilinear grid points. The sum of these three components is expected to be close to the total transport (up to a small residual e.g. due to sub-grid mixing, temporal averaging or the choice of reference value).

\begin{figure}
    \centering
\noindent\includegraphics[width=0.8\textwidth]{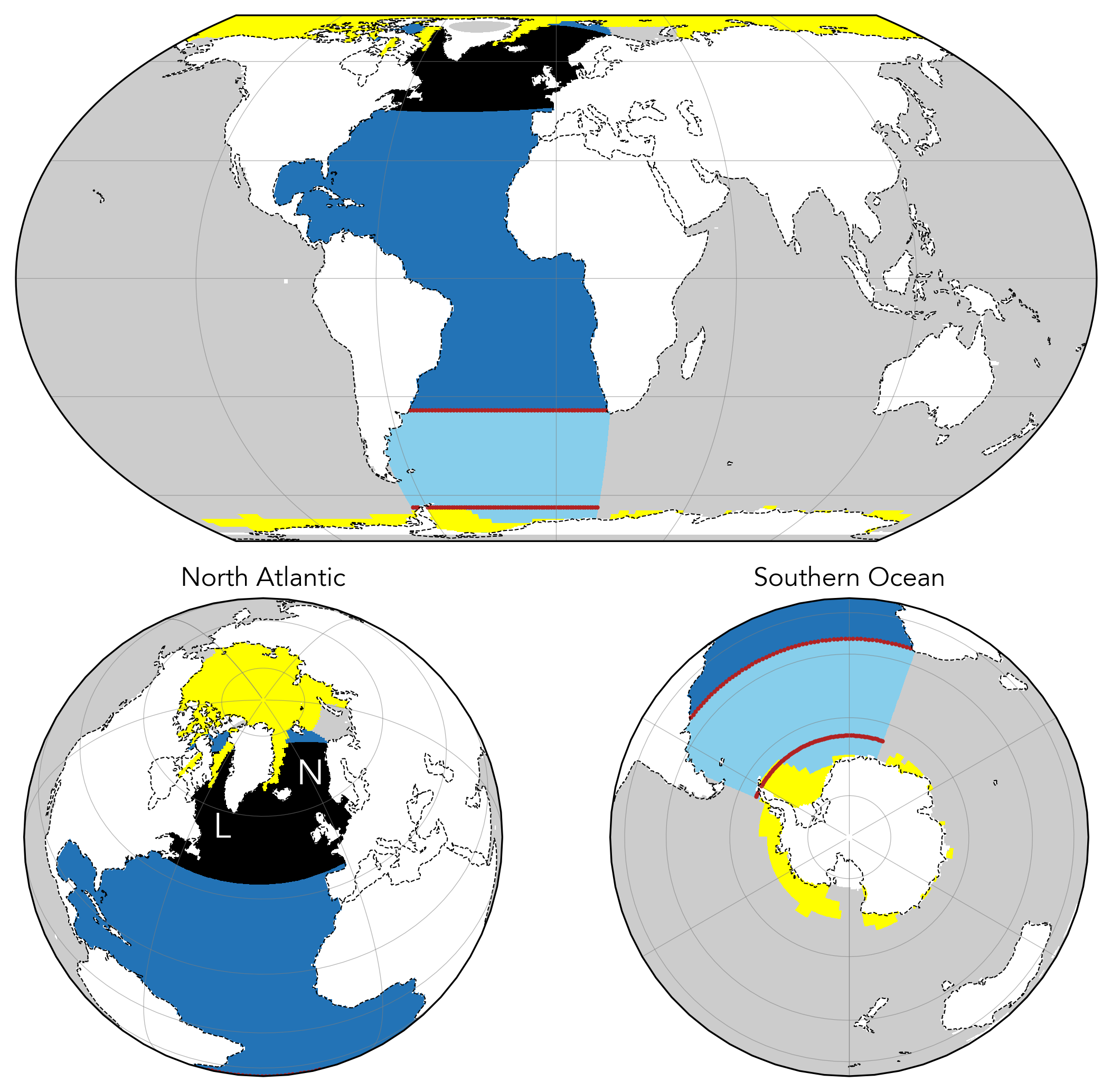}
\caption{Region map of the POP2 model grid, showing the region $\Omega$ (black), the Atlantic basin (blue) and its sector extension in the Southern Ocean (lighter blue). Red zonal lines mark the boundaries of the Southern ocean sector region used to calculate the meridional density gradient $\Delta \rho_y$. The sea ice extent (inferred from a vanishing downward shortwave radiative flux) is shown in yellow. The locations of the Labrador (L) and Nordic (N) Seas are indicated.}
\label{fig:1}
\end{figure}

\begin{figure}
\noindent\includegraphics[width=\textwidth]{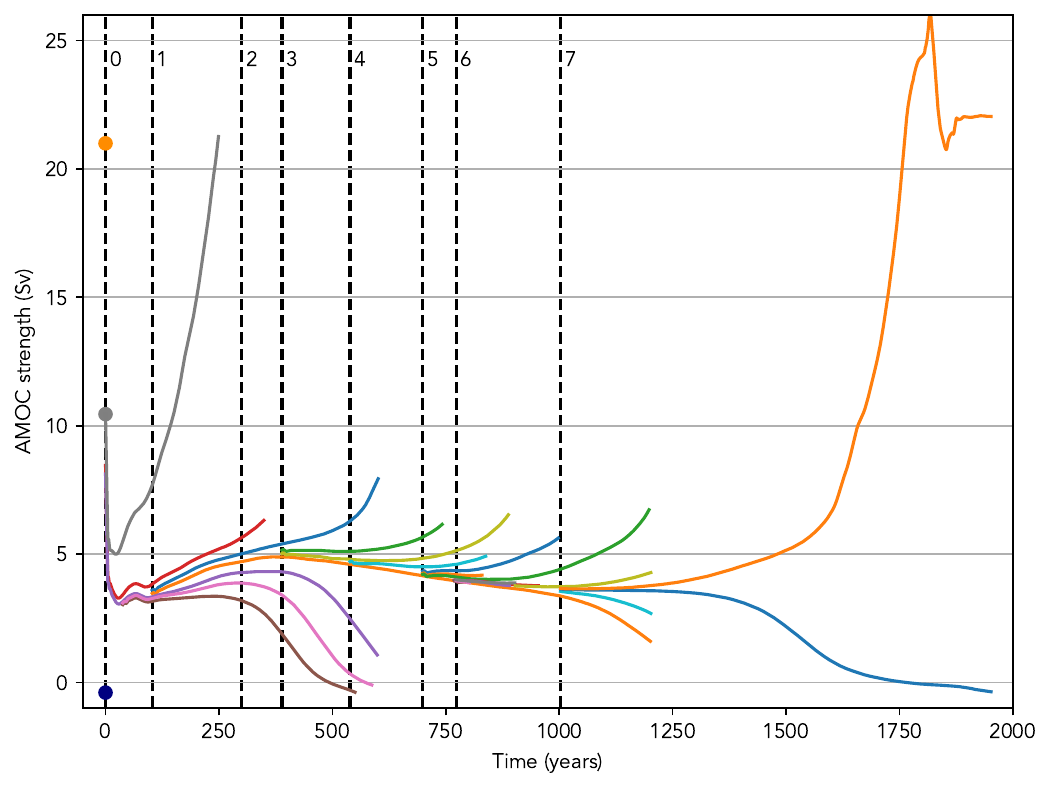}
\caption{All model trajectories generated during the edge tracking procedure, projected onto the AMOC strength at 26$^\circ$N and 1000~m depth. The blue and orange dot at year zero indicate the starting points; the gray dot marks the AMOC strength of the first bisected restart file. Vertical lines (black dashed) indicate re-bisection steps, numbered by iteration.}
\label{fig:s2}
\end{figure}

\begin{figure}
\noindent\includegraphics[width=\textwidth]{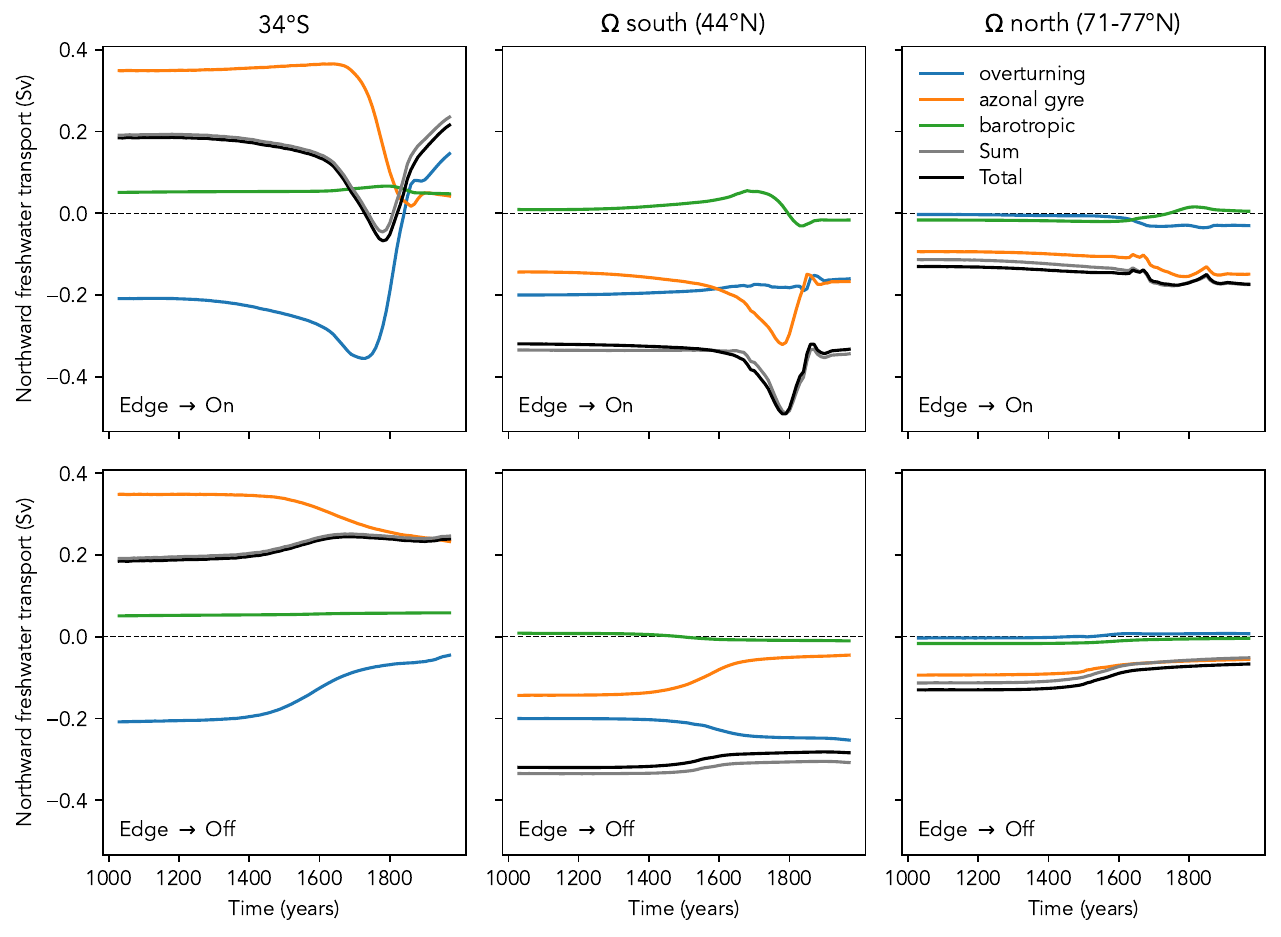}
\caption{Freshwater transport decomposition at different sections (34$^\circ$S, the southern boundary of $\Omega$, and the northern boundary of $\Omega$) along the relaxation trajectory from the edge state to the On (top) and Off (bottom) states. The sum (gray) of the overturning (ov), gyre (az) and barotropic (bt) components is compared to the total transport (black).}
\label{fig:s3}
\end{figure}

\begin{figure}
\noindent\includegraphics[width=\textwidth]{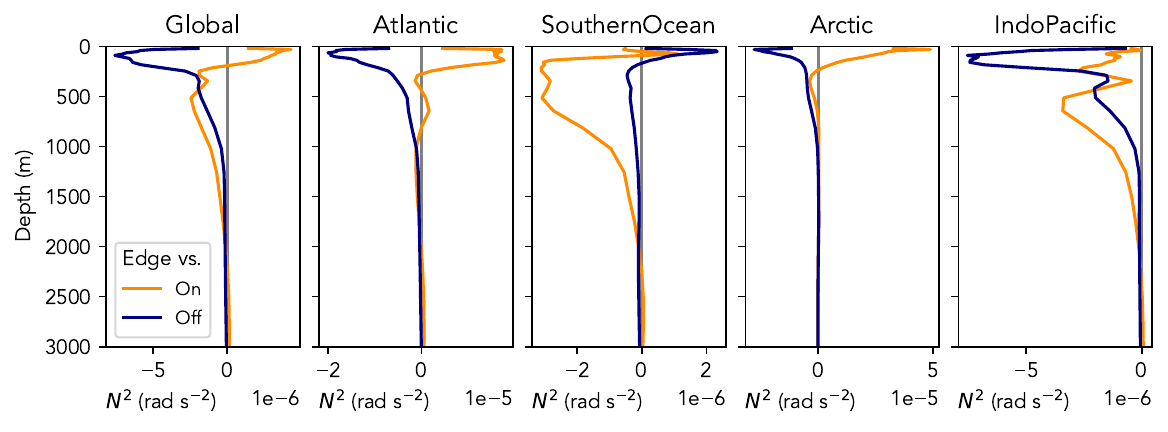}
\caption{Stratification anomaly of the edge state relative to the On (orange) and Off (blue) states, shown as vertical profiles of differences of the Brunt-Väisälä frequency $N^2$, averaged over different ocean basins. Negative values indicate that the edge state is less stratified.}
\label{fig:s4}
\end{figure}

\end{document}